\documentclass[]{biometrika}  

\usepackage{bm}
\usepackage{soul}
\usepackage{amsmath,amssymb}
\usepackage{natbib}
\usepackage{subfigure}
\usepackage{graphics}
\usepackage{epsfig}
\usepackage{multirow}
\usepackage{color}

\def\T{{ \mathrm{\scriptscriptstyle T} }}

\begin{document}




\title{Posterior consistency in linear models under shrinkage priors}

\author{A. ARMAGAN}
\affil{SAS Institute Inc., Cary, North Carolina 27513, USA \email{artin.armagan@sas.com}}

\author{\and D. B. DUNSON}
\affil{Department of Statistical Science, Duke University, Durham, North Carolina 27708, USA \email{dunson@stat.duke.edu}}

\author{\and J. LEE}
\affil{Department of Statistics, Seoul National University, Seoul, 151-747, Korea \email{leejyc@gmail.com}}

\author{\and W. U. BAJWA}
\affil{Department of Electrical and Computer Engineering, Rutgers University, Piscataway, New Jersey 08854, USA \email{waheed.bajwa@rutgers.edu}}

\author{\and N. STRAWN}
\affil{Department of Mathematics, Duke University, Durham, North Carolina 27708, USA \email{nstrawn@math.duke.edu}}

\maketitle

\begin{abstract}
We investigate the asymptotic behavior of posterior distributions of regression coefficients in high-dimensional linear models as the number of dimensions grows with the number of observations.  We show that the posterior distribution concentrates in neighborhoods of the true parameter under simple sufficient conditions. These conditions hold under popular shrinkage priors given some sparsity assumptions.
\end{abstract}


\section{Introduction}
Consider the linear model ${y}_{n}={X}_{n}\beta_{n}^{0}+{\varepsilon}_{n}$, 
where ${y}_{n}$ is an $n$-dimensional vector of responses, ${X}_{n}$ is the $n\times p_n$ design matrix, $\varepsilon_{n} \sim \small{\mbox{N}}\left({0},\sigma^{2}{I}_{n}\right)$ with known $\sigma^2$, and some of the components of $\beta_{n}^{0}$ are zero. Let $\mathcal{A}_n=\{j:\beta_{nj}^{0}\neq 0,j=1,\ldots,p_n\}$ and $|\mathcal{A}_n|=q_n$ denote the set of indices and number of nonzero elements in $\beta_{n}^{0}$.  

In studying the behavior of regression methods in high-dimensional settings, it is increasingly common to allow the number of candidate predictors $p_n$ to grow with sample size $n$.  This is realistic in many applications. In genomics the number of predictors tends to be larger by design for studies with more subjects. In collecting single nucleotide polymorphisms, gene expression, proteomics and so on, one can obtain an immense number of candidate predictors.  However, when $n$ is small, attempting to measure and include all such predictors in the statistical analysis seems unreasonable, so that one tends to collect and analyze increasing subsets of an effectively unbounded number of candidate predictors as sample size increases. In such applications, we are often interested in inferences on the model parameters as much as building a predictive model in order to understand the associations between the response and the candidate predictors.

Our setup is not new, and we follow \cite{ghosal1999} who also focused on asymptotic properties of the posterior on the regression coefficients assuming known $\sigma^2$ and growing $p_n$.  The increasing $p_n$ paradigm induces some challenges relative to the traditional literature on posterior consistency in that growing dimension of $\beta_n^0$ results in a changing $\ell_2$ neighborhood around $\beta_n^0$. This makes it more challenging to show that the posterior assigns all such neighborhoods probability converging to one.  One way to bypass this issue is to focus on the predictive distribution of $y_n$ given $X_n$ as in \cite{jiang2007}.  However, this does not address the common interest in inferences on the regression coefficients. \cite{ghosal1999} and \cite{bontemps2011} provide results on asymptotic normality of the posteriors in linear models for $p_n^4\log p_n=o(n)$ and $p_n\leq n$, respectively. As a corollary, \cite{ghosal1999} states posterior consistency results in linear models when $p_n^3\log n/n\rightarrow 0$ under the usual assumptions on $X_n$. However, both \cite{ghosal1999} and \cite{bontemps2011} require Lipschitz conditions ensuring that the prior is sufficiently flat in a neighborhood of the true $\beta_n^0$. Such conditions are restrictive when using shrinkage priors that are designed to concentrate on sparse $\beta_n$ vectors.

Our main contribution is providing a simple sufficient condition on the prior concentration to achieve the desired asymptotic posterior behavior when $p_n=o(n)$.  Our particular focus is on shrinkage priors, including the Laplace, Student's $t$, generalized double Pareto, and horseshoe-type priors \citep{johnstone2004,carvalho2010,armagan11c,armagan2013}.  There is a rich methodological and applied literature supporting such priors but a lack of theoretical results. 

\section{Sufficient Conditions for Posterior Consistency}
Our results on posterior consistency rely on the following assumptions as $n\rightarrow \infty$:
\begin{enumerate}
\item[](A1) Let $p_n = o(n)$;
\item[](A2) Let $\Lambda_{n\min}$ and $\Lambda_{n\max}$ be the smallest and the largest singular values of $X_n$, respectively. Then  $0< \Lambda_{\min}<\lim\inf_{n\rightarrow\infty}\Lambda_{n\min}/\surd{n}\leq\lim\sup_{n\rightarrow\infty}\Lambda_{n\max}/\surd{n}<\Lambda_{\max}<\infty$;
\item[](A3) Let $\sup_{j=1,\ldots,p_n}|\beta_{nj}^{0}|<\infty$;
\item[](A4) Let $q_n =o\{n^{1-\rho/2}/(\surd p_n\log n)\}$ for $\rho\in(0,2)$;
\item[](A5) Let $q_n = o(n/\log n)$. 
\end{enumerate}
Assumptions (A4) and (A5) will be used in different settings. 

\begin{lemma}
Let $\mathcal{B}_{n}:=\{\beta_{n}:\|\beta_{n}-\beta_{n}^{0}\|>\epsilon\}$ where $\epsilon>0$. To test $H_0: \beta_{n}=\beta_{n}^{0}$ vs $H_1: \beta_{n}\in \mathcal{B}_{n}$, we define a test function $\Phi_{n}(y_{n})=I(y_{n}\in \mathcal{C}_{n})$ where the critical region is $\mathcal{C}_{n}:=\{y_{n}:\|\hat{\beta}_{n}-\beta_{n}^{0}\|>\epsilon/2\}$ and $\hat{\beta}_n = (X_n^TX_n)^{-1}X_n^Ty_n$.  Then, under assumptions (A1) and (A2), as $n\rightarrow \infty$,
\begin{enumerate}
\item $E_{\beta_{n}^{0}}(\Phi_{n})\leq \exp\{-\epsilon^2n\Lambda^2_{\min}/(16\sigma^2)\}$,
\item $\sup_{\beta_{n}\in \mathcal{B}_{n}}E_{\beta_{n}}(1-\Phi_{n})\leq \exp\{-\epsilon^2n\Lambda^2_{\min}/(16\sigma^2)\}$.
\end{enumerate}
\end{lemma}

\begin{theorem} Given Lemma 1, the posterior of $\beta_n$ under prior $\Pi_{n}(\beta_n)$ is strongly consistent, that is, for any $\epsilon>0$, $\Pi_n(\mathcal{B}_n |y_n)=\Pi_n(\beta_n : ||\beta_n - \beta_{n}^{0} || > \epsilon |y_n) \rightarrow 0$ $\mbox{pr}_{\beta^0_n}$--almost surely as $n\rightarrow \infty$, if
\begin{equation}
 \Pi_{n}\left(\beta_{n}:\|\beta_{n}-\beta_{n}^{0}\|<\frac{\Delta}{n^{\rho/2}}\right)>\exp(-dn)\nonumber
\end{equation}
for all $0<\Delta<\epsilon^2\Lambda^2_{\min}/(48\Lambda^2_{\max})$ and $0<d<\epsilon^2\Lambda^2_{\min}/(32\sigma^2)-3\Delta\Lambda^2_{\max}/(2\sigma^2)$ and some $\rho>0$.
\end{theorem}

Theorem 1 provides a simple sufficient condition on the concentration of the prior around sparse $\beta_n^0$. We use Theorem 1 to provide conditions on $\beta_n^0$ under which specific shrinkage priors achieve posterior consistency focusing on priors that assume independent and identically distributed elements of $\beta_n$.  

\subsection{Laplace Prior}

\begin{theorem} Under assumptions (A1)--(A4), the Laplace prior $f(\beta_{nj}|s_n)=(1/2s_n)\exp(-|\beta_{nj}|/s_n)$
with scale parameter $s_n$ yields a strongly consistent posterior if $s_n=  C/(\surd p_n n^{\rho/2}\log n)$ for finite $C>0$.
\end{theorem}

\subsection{Student's $t$ Prior}

The density function for the scaled Student's $t$ distribution is 
\begin{equation} 
f(\beta_{j}|s,d_0)=\frac{1}{s\surd d_0 \mbox{B}(1/2,d_0/2)}\left(1+\frac{\beta_j^2}{s^2 d_0}\right)^{-(d_0+1)/2}\nonumber,
\end{equation}
with scale $s$, degrees of freedom $d_0$, and $\mbox{B}(\cdot)$ denoting the beta function.

\begin{theorem} Under assumptions (A1)--(A3) and (A5), the scaled Student's $t$ prior with parameters $s_n$ and $d_{0n}$ yields a strongly consistent posterior if $d_{0n}=d_0\in (2,\infty)$ and $s_n = C/(\surd p_n n^{\rho/2}\log n)$  for finite $\rho>0$ and $C>0$.
\end{theorem}

\subsection{Generalized Double Pareto Prior}

As defined by \cite{armagan2013}, the generalized double Pareto density is given by
\begin{equation}
f(\beta_j|\alpha,\eta)=\frac{\alpha}{2\eta}\left(1+\frac{|\beta_j|}{\eta}\right)^{-(\alpha+1)}, \ \ \alpha,\eta>0. \nonumber
\end{equation}

\begin{theorem} Under assumptions (A1)--(A3) and (A5), the generalized double Pareto prior with parameters $\alpha_n$ and $\eta_n$ yields a strongly consistent posterior if $\alpha_n=\alpha\in(2,\infty)$ and $\eta_n = C/(\surd p_n n^{\rho/2}\log n)$ for finite $\rho>0$ and $C>0$.
\end{theorem}

\subsection{Horseshoe-like Priors}

As defined in \cite{armagan11c}, generalized beta scale mixtures of normals are obtained by the following three equivalent representations:
\begin{eqnarray}
\beta_j&\sim& \small{\mbox{N}}(0,1/\varrho_j-1), f(\varrho_j)= \frac{\Gamma(a_0+b_0)}{\Gamma(a_0)\Gamma(b_0)}\xi^{b_0}\varrho_j^{b_0-1}(1-\varrho_j)^{a_0-1}\left\{1+(\xi-1)\varrho_j\right\}^{-(a_0+b_0)} \label{rep1}\\
\beta_{j}&\sim& \small{\mbox{N}}(0,\tau_{j}), \tau_{j} \sim {\mbox{Ga}}(a_0,\lambda_{j}), \lambda_{j} \sim \small{\mbox{Ga}}(b_0,\xi)\nonumber\\
\beta_{j}&\sim& \small{\mbox{N}}(0,\tau_{j}), f(\tau_{j})=\frac{\Gamma(a_0+b_0)}{\Gamma(a_0)\Gamma(b_0)}\xi^{-a_0}\tau^{a_0-1}(1+\tau_{j}/\xi)^{-(a_0+b_0)}\nonumber
\end{eqnarray}
where $a_0,b_0,\xi>0$. Due to the representation in (\ref{rep1}) and the work by \cite{carvalho2010}, we refer to these priors as \emph{horseshoe-like}. The above formulation yields a general family that covers special cases discussed in \cite{johnstone2004}, a technical report by Griffin \& Brown (2007) and \cite{carvalho2010}. The resulting marginal density on $\beta_j$ is 
\begin{equation}
f(\beta_j|a_0,b_0,\xi)=\frac{\Gamma(b_0+1/2)\Gamma(a_0+b_0){\mbox{U}}\{b_0+1/2,3/2-a_0,\beta_{j}^2/(2\xi)\}}{(2\pi\xi)^{1/2}\Gamma(a_0)\Gamma(b_0)},
\label{marginal}
\end{equation}
where ${\mbox{U}}(\cdot)$ denotes the confluent hypergeometric function of the second kind. 
\begin{theorem} Under assumptions (A1)--(A3) and (A5), the prior in (\ref{marginal}) with parameters $a_{0n}=a_0\in(0,\infty)$, $b_{0n}=b_0\in(1,\infty)$ and $\xi_n$ yields a strongly consistent posterior if $\xi_n = C/(p_n n^{\rho}\log n)$ for finite $\rho>0$ and $C>0$.
\end{theorem}
 
\section{Final Remarks}

Our analysis is heavily dependent on the construction of good tests.  Results can be extended utilizing appropriate tests relying on an estimator with asymptotically vanishing probability of being outside of a \emph{shrinking} neighborhood of the truth.  For instance, one could use results similar to 
\cite{bickel2009} given additional conditions on $X_n$.  Theorem 7.2 of \cite{bickel2009} states that
\begin{equation}
\mbox{pr}_{\beta_{n}^{0}}\left(\|\hat{\beta}_{nL}-\beta_n^{0}\|_2^2> M\frac{a_n\log p_n}{n}\right)\leq p_n^{1-a_n^2/8}
\label{lasso}
\end{equation}
for $a_n>2\surd2$ and for some $M>0$, where $\hat{\beta}_{nL}$ denotes the Lasso estimator. Hence using (\ref{lasso}), in a similar fashion to Lemma 1, we can obtain consistent tests with an $\epsilon$-neighborhood contracting at a rate $\mathcal{O}\{(a_n\log p_n)^{1/2} /\surd n\}$. Assuming $q_n<\infty$ for simplicity and letting $a_n=\mathcal{O}(\log n)$, following Theorems 1, 3, 4 and 5, we anticipate that under the Student's $t$, generalized double Pareto and horseshoe-like priors, a \emph{near-optimal} contraction rate of $\mathcal{O}\{(\log n \log p_n)^{1/2} /\surd n\}$ is possible. 

As in almost all of the Bayesian asymptotic literature, we have focused on sufficient conditions.  Our conditions are practically appealing in allowing priors to be screened for their usefulness in high-dimensional settings.  However, it would be of substantial interest to additionally provide theory allowing one to rule out the use of certain classes of priors in particular settings.  

\section{Technical Details}
\begin{proof}[of Lemma 1] Noting that $\hat{\beta}_{n}=(X_{n}^\T X_{n})^{-1}X_{n}^\T y_{n}$, $E_{\beta_{n}^{0}}(\Phi_{n})=\mbox{pr}_{\beta_{n}^{0}}(\|\hat{\beta}_n-\beta_n^0\|>\epsilon/2)\leq\mbox{pr}_{\beta_{n}^{0}}\{\chi^{2}_{p_n} >\epsilon^2n\Lambda_{\min}^2/(4\sigma^{2})\}$ where $\chi_{p}^2$ is a chi-squared distributed random variable with $p$ degrees of freedom. The inequality is attained using assumption (A2). Similarly, $\sup_{\beta_{n}\in \mathcal{B}_n}E_{\beta_{n}}(1-\Phi_{n})\leq\sup_{\beta_{n}\in \mathcal{B}_n}\mbox{pr}_{\beta_{n}}(|\|\hat{\beta}_{n}-\beta_{n}\|-\|\beta_{n}^{0}-\beta_{n}\||\leq \epsilon/2)\leq\sup_{\beta_{n}\in \mathcal{B}_n}\mbox{pr}_{\beta_{n}}(\|\hat{\beta}_{n}-\beta_{n}\|\geq -\epsilon/2+\|\beta_{n}^{0}-\beta_{n}\|)=\mbox{pr}_{\beta_{n}}(\|\hat{\beta}_{n}-\beta_{n}\|\geq \epsilon/2)\leq \mbox{pr}_{\beta_{n}^{0}}\{\chi^{2}_{p_n} >\epsilon^2n\Lambda_{\min}^2/(4\sigma^{2})\}$. Simplifying the inequality $\mbox{pr}\{\chi^2_{p}-p\geq2(px)^{1/2}+2x\}\leq \exp(-x)$ by \cite{laurent2000}, we state that $\mbox{pr}(\chi^2_p\geq x)\leq\exp(-x/4)$ if $x\geq 8p$. Then, using assumption (A1), as $n\rightarrow \infty$, 
\begin{eqnarray}
E_{\beta_{n}^{0}}(\Phi_{n})&\leq&\exp\{-\epsilon^2n\Lambda_{\min}^2/(16\sigma^2)\},\nonumber\\
\label{eq1}
\sup_{\beta_{n}\in \mathcal{B}_n}E_{\beta_{n}}(1-\Phi_{n})&\leq&\exp\{-\epsilon^2n\Lambda_{\min}^2/(16\sigma^2)\}.\nonumber
\label{eq2}
\end{eqnarray}
This completes the proof.
\end{proof}

\begin{proof}[of Theorem 1]
Our proof relies on a technique originally devised by \cite{sch}. The posterior probability of $\mathcal{B}_{n}$ is given by
\begin{eqnarray}
\Pi_{n}(\mathcal{B}_{n}|y_{n})&=&\frac{\int_{\mathcal{B}_{n}}\{f(y_{n}|\beta_{n})/f(y_{n}|\beta_{n}^{0})\}\Pi(d\beta_{n})}{\int\{f(y_{n}|\beta_{n})/f(y_{n}|\beta_{n}^{0})\}\Pi(d\beta_{n})}\nonumber\\
&\leq&\Phi_{n}+\frac{(1-\Phi_{n})J_{\mathcal{B}_{n}}}{J_{n}}\nonumber\\
&=&I_{1}+I_{2}/J_{n},\nonumber
\end{eqnarray}
\label{eq4a}where $J_{\mathcal{B}_{n}}=\int_{\mathcal{B}_{n}} \{f(y_n|\beta_n)/f(y_n|\beta_n^0)\}\Pi(d\beta_n)$ and $J_n=J_{\Re^{p_n}}$. We need to show that $I_{1}+I_{2}/J_{n}\rightarrow 0$ $\mbox{pr}_{\beta^0_n}$--almost surely as $n\rightarrow \infty$. Let $b=\epsilon^2\Lambda^2_{\min}/(16\sigma^2)$. For sufficiently large $n$, $\mbox{pr}_{\beta^{0}_{n}}\{I_{1}\geq \exp(-bn/2)\}\leq\exp(bn/2)E_{\beta_{n}^{0}}(I_{1})=\exp(-bn/2)$ using Lemma 1. This implies that $\sum_{n=1}^{\infty}\mbox{pr}_{\beta^{0}_{n}}\{I_{1}\geq \exp(-bn/2)\}<\infty$ and hence by the Borel--Cantelli lemma $\mbox{pr}_{\beta_{0}}\{I_{1}\geq \exp(-bn/2) \ \ \mbox{infinitely often}\}=0$. We next look at the behavior of $I_{2}$:
\begin{eqnarray}
E_{\beta_{n}^{0}}(I_{2})&=&E_{\beta_{n}^{0}}\{(1-\Phi_{n})J_{\mathcal{B}_{n}}\}\nonumber\\
&=&E_{\beta_{n}^{0}}\left\{(1-\Phi_{n})\int_{\mathcal{B}_{n}}\frac{f(y_{n}|\beta_{n})}{f(y_{n}|\beta_{n}^{0})}\Pi_{n}(d\beta_{n})\right\}\nonumber\\
&=&\int_{\mathcal{B}_{n}}\int(1-\Phi_{n})f(y_{n}|\beta_{n})dy_{n}\Pi_{n}(d\beta_{n})\nonumber\\
&\leq&\Pi_{n}(\mathcal{B}_{n})\sup_{\beta_{n}\in \mathcal{B}_{n}}E_{\beta_{n}}(1-\Phi_{n})\nonumber\\
&\leq&\exp(-bn)\nonumber
\end{eqnarray}
Then for sufficiently large $n$, $\mbox{pr}_{\beta^{0}_{n}}\{I_{2}\geq \exp(-bn/2)\}\leq \exp(-bn/2)$ using Lemma 1. Again $\sum_{n=1}^{\infty}\mbox{pr}_{\beta^{0}_{n}}\{I_{2}\geq \exp(-bn/2)\}<\infty$ and hence by the Borel--Cantelli lemma $\mbox{pr}_{\beta_{0}}\{I_{2}\geq \exp(-bn/2) \ \ \mbox{infinitely often}\}=0$. 

We have shown that both $I_{1}$ and $I_{2}$ tend towards zero exponentially fast. Now we analyze the behavior of $J_{n}$. To complete the proof, we need to show that $\exp(bn/2) J_{n}\rightarrow \infty$ $\mbox{pr}_{\beta^0_n}$--almost surely as $n\rightarrow \infty$. 
\begin{eqnarray}
\exp(b n/2)J_{n}&=&\exp(b n/2) \int \exp\left\{-n\frac{1}{n}\log \frac{f(y_{n}|\beta_{n}^{0})}{f(y_{n}|\beta_{n})}\right\}\Pi_{n}(d\beta_{n})\nonumber\\
&\geq& \exp\{(b/2-\nu)n\}\Pi_{n}(\mathcal{D}_{n,\nu})
\label{eq5}
\end{eqnarray}
where $\mathcal{D}_{n,\nu}=\{\beta_{n}:n^{-1}\log\{f(y_{n}|\beta_{n}^{0})/f(y_{n}|\beta_{n})\}<\nu\}=\{\beta_{n}:n^{-1}(\|y_{n}-X_{n}\beta_{n}\|^2-\|y_{n}-X_{n}\beta_{n}^{0}\|^2)<2\sigma^{2}\nu\}$ for any $0<\nu<b/2$. Then $\Pi_{n}(\mathcal{D}_{n,\nu})\geq\Pi_{n}\{\beta_{n}:n^{-1}|\|y_{n}-X_{n}\beta_{n}\|^2-\|y_{n}-X_{n}\beta_{n}^{0}\|^2|<2\sigma^{2}\nu\}$. Using the identity $x^2-x_0^2=2x_0(x-x_0)+(x-x_0)^2$ for all $x,x_0\in \Re$,
\begin{eqnarray}
\Pi_{n}(\mathcal{D}_{n,\nu})&\geq&\Pi_{n}\left\{\beta_{n}:n^{-1}\left|2\|y_{n}-X_{n}\beta_{n}^{0}\|(\|y_{n}-X_{n}\beta_{n}\|-\|y_{n}-X_{n}\beta_{n}^{0}\|)\right.\right.\nonumber\\
& &+\left.\left.(\|y_{n}-X_{n}\beta_{n}\|-\|y_{n}-X_{n}\beta_{n}^{0}\|)^2\right|<2\sigma^{2}\nu\right\}\nonumber\\
&\geq&\Pi_{n}\left\{\beta_{n}:n^{-1}(2\|y_{n}-X_{n}\beta_{n}^{0}\|\|X_{n}\beta_{n}-X_{n}\beta_{n}^{0}\|+\|X_{n}\beta_{n}-X_{n}\beta_{n}^{0}\|^2)<2\sigma^{2}\nu\right\}\nonumber\\
&\geq&\Pi_{n}\left(\beta_{n}:n^{-1}\|X_{n}\beta_{n}-X_{n}\beta_{n}^{0}\|<\frac{2\sigma^{2}\nu}{3\kappa_n},\|X_{n}\beta_{n}-X_{n}\beta_{n}^{0}\|< \kappa_n \right)
\label{eq7}
\end{eqnarray}
given that $\|y_{n}-X_{n}\beta_{n}^{0}\|\leq \kappa_{n}$. For $\kappa_{n}=n^{(1+\rho)/2}$ with $\rho>0$ and $\kappa_n^2/\sigma^2\geq 8n$, $\mbox{pr}_{\beta_{n}^{0}}(y_{n}:\|y_{n}-X_{n}\beta_{n}^{0}\|^2> \kappa_{n}^2)=\mbox{pr}_{\beta_{n}^{0}}(y_{n}:\chi^{2}_{n}>\kappa_{n}^2/\sigma^2)\leq\exp\{-\kappa_{n}^2/(4\sigma^2)\}$.
Since $\sum_{n=1}^{\infty}\mbox{pr}_{\beta_{n}^{0}}(y_{n}:\|y_{n}-X_{n}\beta_{n}^{0}\|> \kappa_{n})<\infty$, by the Borel--Cantelli lemma  $\mbox{pr}_{\beta_{n}^{0}}(y_{n}:\|y_{n}-X_{n}\beta_{n}^{0}\|> \kappa_{n} \ \ \mbox{infinitely often})=0$. Following from (\ref{eq7}) and the fact that $\kappa_n \rightarrow \infty$, as $n \rightarrow \infty$, for sufficiently large $n$, $\Pi_{n}(\mathcal{D}_{n,\nu})\geq\Pi_{n}\{\beta_{n}:n^{-1}\|X_{n}\beta_{n}-X_{n}\beta_{n}^{0}\|<2\sigma^{2}\nu/(3\kappa_n)\}\geq\Pi_{n}(\beta_{n}:\|\beta_{n}-\beta_{n}^{0}\|<\Delta/n^{\rho/2})$, where $\Delta=2\sigma^2\nu/(3\Lambda_{\max})$. Hence following (\ref{eq5}), $\Pi_{n}(\mathcal{B}_{n}|y_{n})\rightarrow 0$ $\mbox{pr}_{\beta^0_n}$--almost surely as $n\rightarrow \infty$ if $\Pi_{n}(\beta_{n}:\|\beta_{n}-\beta_{n}^{0}\|<\Delta/n^{\rho/2})>\exp(-dn)$ for all $0<d<b/2-\nu$. This completes the proof.
\end{proof}
\begin{proof}[of Theorem 2] We need to calculate the probability assigned to the region $\{\beta_n:\|\beta_{n}-\beta_{n}^{0}\|<\Delta/n^{\rho/2}\}$ under the Laplace prior.
\begin{eqnarray}
& &\Pi_{n}\left(\beta_{n}:\|\beta_{n}-\beta_{n}^{0}\|<\frac{\Delta}{n^{\rho/2}}\right) = \Pi_{n}\left\{\beta_{n}:\sum_{j\in \mathcal{A}_n}(\beta_{nj}-\beta_{nj}^{0})^2+\sum_{j\notin \mathcal{A}_n}\beta_{nj}^2<\frac{\Delta^2}{n^{\rho}}\right\} \nonumber\\
& &\hspace{20pt}\geq\prod_{j\in\mathcal{A}_n}\left\{\Pi_{n}\left(\beta_{nj}:|\beta_{nj}-\beta_{nj}^{0}|<\frac{\Delta}{\surd p_n n^{\rho/2}}\right)\right\}\nonumber\\
& &\hspace{40pt}\times\Pi_{n}\left\{\beta_{n}^{j\notin\mathcal{A}}:\sum_{j\notin \mathcal{A}_n}\beta_{nj}^2<\frac{(p_n-q_n)\Delta^2}{p_n n^{\rho}}\right\}\nonumber\\
& &\hspace{20pt}\geq\prod_{j\in\mathcal{A}_n}\left\{\Pi_{n}\left(\beta_{nj}:|\beta_{nj}-\beta_{nj}^{0}|<\frac{\Delta}{\surd p_n n^{\rho/2}}\right)\right\}\left\{1-\frac{p_n n^{\rho}E\left(\sum_{j\notin\mathcal{A}_n}\beta_{nj}^2\right)}{(p_n-q_n)\Delta^2}\right\}
\label{lap1}
\end{eqnarray}
where $E(\beta_{nj}^2)$ can verified to be $2s_n^2$. Following from (\ref{lap1})
\begin{eqnarray}
& &\Pi_{n}\left(\beta_{n}:\|\beta_{n}-\beta_{n}^{0}\|<\frac{\Delta}{n^{\rho/2}}\right) \geq \nonumber\\
& &\hspace{20pt}\left\{\frac{\Delta}{\surd p_n n^{\rho/2}s_n}\exp\left(-\frac{\sup_{j\in\mathcal{A}_n}|\beta_{nj}^{0}|}{s_n}-\frac{\Delta}{s_n \surd p_n n^{\rho/2}}\right)\right\}^{q_n}\left(1-\frac{2p_n n^{\rho}s_n^2}{\Delta^2}\right).
\label{lap2}
\end{eqnarray}
Taking the negative logarithm of both sides of (\ref{lap2}) and letting $s_n = C/(\surd p_n n^{\rho/2}\log n)$ for some $C>0$, we obtain
\begin{eqnarray}
& &-\log \Pi_{n}\left(\beta_{n}:\|\beta_{n}-\beta_{n}^{0}\|<\frac{\Delta}{n^{\rho/2}}\right) \leq -q_n \log \Delta+q_n\log C-q_n\log\log n\nonumber\\
& & \hspace{20pt} -\log\left\{1-\frac{2C^2}{\Delta^2 (\log n)^2}\right\} +\frac{q_n\Delta\log n}{C}+\frac{q_n\surd p_n n^{\rho/2}\log n \sup_{j\in\mathcal{A}_n}|\beta_{nj}^{0}|}{C}
\label{lap3}
\end{eqnarray}
as $n\rightarrow \infty$. It is easy to see that the dominating term in (\ref{lap3}) is the last one and $-\log \Pi_{n}(\beta_{n}:\|\beta_{n}-\beta_{n}^{0}\|<\Delta/ n^{\rho/2})<dn$ for all $d>0$. This completes the proof.
\end{proof}

\begin{proof}[of Theorem 3] 
$E(\beta_{nj}^2)$, in this case, is given by $d_{0} s_n^2/(d_{0}-2)$. For the sake of simplicity, we let $d_{0}=3$. Then following from (\ref{lap1}) 
\begin{eqnarray}
& &\Pi_{n}\left(\beta_{n}:\|\beta_{n}-\beta_{n}^{0}\|<\frac{\Delta}{n^{\rho/2}}\right) \geq \left(1-\frac{3p_n n^{\rho}s_n^2}{\Delta^2}\right)\nonumber\\
& &\hspace{20pt}\times\left[\frac{2\Delta}{\surd p_n n^{\rho/2}s_n\surd 3\mbox{B}(1/2,3/2)}\left\{1+\frac{2\sup_{j\in\mathcal{A}_n}(\beta_{nj}^{0})^2}{3s_n^2}+\frac{2\Delta^2}{3s_n^2 p_n n^{\rho}}\right\}^{-2}\right]^{q_n}.
\label{t2}
\end{eqnarray}
Taking the negative logarithm of both sides of (\ref{t2}) and letting $s_n =C/(\surd p_n n^{\rho/2}\log n)$ for some $C>0$, we obtain
\begin{eqnarray}
& &-\log \Pi_{n}\left(\beta_{n}:\|\beta_{n}-\beta_{n}^{0}\|<\frac{\Delta}{n^{\rho/2}}\right) \leq q_n\log \left\{\frac{\surd 3 C\mbox{B}(1/2,3/2)}{2\Delta}\right\}-q_n\log\log n\nonumber\\
& & \hspace{20pt} -\log\left\{1-\frac{C^2}{\Delta^2 (\log n)^2}\right\} +2q_n\log\left\{1+\frac{2p_n n^{\rho}\log n\sup_{j\in\mathcal{A}_n}(\beta_{nj}^{0})^2}{3C^2}+\frac{2\Delta^2(\log n)^2}{3C^2}\right\} \nonumber\\
\label{t3}
\end{eqnarray}
as $n\rightarrow \infty$. It is easy to see that the dominating term in (\ref{t3}) is the last one and $-\log \Pi_{n}(\beta_{n}:\|\beta_{n}-\beta_{n}^{0}\|<\Delta/ n^{\rho/2})<dn$ for all $d>0$. The result can be easily shown to hold for all $d_0\in(2,\infty)$. This completes the proof.
\end{proof}

\begin{proof}[of Theorem 4] 
$E(\beta_{nj}^2)$, in this case, can verified to be $2\eta_n^2/(\alpha^2-3\alpha+2)$ for $\alpha>2$. For the sake of simplicity, we let $\alpha=3$. Then following from (\ref{lap1}) 
\begin{eqnarray}
& &\Pi_{n}\left(\beta_{n}:\|\beta_{n}-\beta_{n}^{0}\|<\frac{\Delta}{n^{\rho/2}}\right) \geq \nonumber\\
& &\hspace{20pt}\left\{\frac{3\Delta}{\surd p_n n^{\rho/2}\eta_n}\left(1+\frac{\sup_{j\in\mathcal{A}_n}|\beta_{nj}^{0}|}{\eta_n}+\frac{\Delta}{\eta_n \surd p_n n^{\rho/2}}\right)^{-4}\right\}^{q_n}\left(1-\frac{p_n n^{\rho}\eta^2}{\Delta^2}\right).
\label{gdp2}
\end{eqnarray}
Taking the negative logarithm of both sides of (\ref{gdp2}) and letting $\eta_n = C/(\surd p_n n^{\rho/2}\log n)$ for some $C>0$, we obtain
\begin{eqnarray}
& &-\log \Pi_{n}\left(\beta_{n}:\|\beta_{n}-\beta_{n}^{0}\|<\frac{\Delta}{n^{\rho/2}}\right) \leq -q_n \log 3\Delta-3q_n\log C-q_n\log\log n\nonumber\\
& & \hspace{20pt} -\log\left\{1-\frac{C^2}{\Delta^2 (\log n)^2}\right\} +4q_n\log\left(C+\Delta\log n+\surd p_n n^{\rho/2}\log n \sup_{j\in\mathcal{A}_n}|\beta_{nj}^{0}|\right)
\label{gdp3}
\end{eqnarray}
as $n\rightarrow \infty$. It is easy to see that the dominating term in (\ref{gdp3}) is the last one and $-\log \Pi_{n}(\beta_{n}:\|\beta_{n}-\beta_{n}^{0}\|<\Delta/ n^{\rho/2})<dn$ for all $d>0$. The result can be easily shown to hold for all $\alpha\in(2,\infty)$. This completes the proof.
\end{proof}

\begin{proof}[of Theorem 5] 
Similarly to the previous cases, we can show that $E(\beta_{nj}^2)=\xi_n\Gamma(a_0+1)\Gamma(b_0-1)/\{\Gamma(a_0)\Gamma(b_0)\}$. Then following from (\ref{lap1}) 
\begin{eqnarray}
& &\Pi_{n}\left(\beta_{n}:\|\beta_{n}-\beta_{n}^{0}\|<\frac{\Delta}{n^{\rho/2}}\right) \geq \left\{1-\frac{p_n n^{\rho}E(\beta_{nj}^2)}{\Delta^2}\right\}\left(\frac{2\Delta}{\surd p_n n^{\rho/2}}\right)^{q_n}\nonumber\\
& &\hspace{20pt}\times\left[\frac{{\mbox{U}}\{b_0+1/2,3/2-a_0,\sup_{j\in\mathcal{A}_n}(\beta_{nj}^0)^2/\xi_n+\Delta/( p_n n^{\rho}\xi_n)\}}{(2\pi\xi_n)^{1/2}\Gamma(a_0)\Gamma(b_0)\Gamma(b_0+1/2)^{-1}\Gamma(a_0+b_0)^{-1}}\right]^{q_n}.
\label{neg2}
\end{eqnarray}
We can use the expansion ${\mbox{U}}(a,b,z)=z^{-a}\{\sum_{m=0}^{R-1}(a)_m(1+a-b)_m(-z)^{m}/m!+\mathcal{O}(|z|^{-R})\}$ for large $z$, where $(a)_m=a(a+1)\ldots(a+m-1)$ and $R$th term is the smallest in the expansion \citep{abramowitz}. Letting $R=1$, 
for sufficiently large $n$, (\ref{neg2}) can be further bounded as
\begin{eqnarray}
&&\Pi_{n}\left(\beta_{n}:\|\beta_{n}-\beta_{n}^{0}\|<\frac{\Delta}{n^{\rho/2}}\right) > \left\{1-\frac{p_n n^{\rho}E(\beta_{nj}^2)}{\Delta^2}\right\}\nonumber\\
&&\hspace{20pt} \times\left[\frac{\surd2\Delta\Gamma(b_0+1/2)\Gamma(a_0+b_0)}{\surd p_n n^{\rho/2}\surd \xi_n\surd\pi\Gamma(a_0)\Gamma(b_0)\{\sup_{j\in\mathcal{A}_n}(\beta_{nj}^0)^2/\xi_n+\Delta/(p_n n^{\rho}\xi_n)\}^{(b_0+1/2)}}\right]^{q_n}.\nonumber\\
\label{neg3}
\end{eqnarray}
Taking the negative logarithm of both sides of (\ref{neg3}) and letting $\xi_n = C/(p_n n^{\rho}\log n)$ for some $C>0$, we obtain
\begin{eqnarray}
& &-\log \Pi_{n}\left(\beta_{n}:\|\beta_{n}-\beta_{n}^{0}\|<\frac{\Delta}{n^{\rho/2}}\right) <\nonumber\\
& &\hspace{20pt}-q_n\log\left\{\frac{\surd 2\Delta\Gamma(b_0+1/2)\Gamma(a_0+b_0)}{\surd C\surd \pi\Gamma(a_0)\Gamma(b_0)}\right\}-\log\left\{1-\frac{C\Gamma(a_0+1)\Gamma(b_0-1)}{\log n \Delta\Gamma(a_0)\Gamma(b_0)}\right\}\nonumber\\
& &\hspace{20pt}-\frac{q_n}{2}\log\log n+q_n\left(b_0+\frac{1}{2}\right)\log \left\{\frac{p_n n^{\rho}\log n \sup_{j\in\mathcal{A}_n}(\beta_{nj}^0)^2}{C}+\frac{\Delta\log n}{C}\right\}
\label{neg4}
\end{eqnarray}
as $n\rightarrow \infty$. It is easy to see that the dominating term in (\ref{neg4}) is the last one and $-\log \Pi_{n}(\beta_{n}:\|\beta_{n}-\beta_{n}^{0}\|<\Delta/ n^{\rho/2})<dn$ for all $d>0$. This completes the proof.
\end{proof}

\section*{Acknowledgements}
This work was supported by the National Institute of Environmental Health
Sciences. The content is solely the responsibility of the authors and does not necessarily represent the official views
of the National Institute of Environmental Health Sciences or the National Institutes of Health. Jaeyong Lee was supported by Advanced Research Center Program (S/ERC), the National Research Foundation of Korea grant funded by the Korean government (MSIP). Waheed U. Bajwa was supported in part by the NSF. Nate Strawn was supported by DARPA Mathematics of Sensing, Exploitation, and Execution (MSEE) program (managed by Dr. Tony Falcone).

\bibliographystyle{biometrika} 
\bibliography{paper_v15} 

\end{document}